# FLEXOELECTRICITY IN ANTIFERROELECTRIC CERAMICS

P. Vales-Castro[1], Krystian Roleder[2], Lei Zhao[3], Jing-Feng Li[3], Dariusz Kajewski[2], Gustau Catalan[1,4]

[1] Catalan Institute of Nanoscience and Nanotechnology (ICN2), Campus Universitat Autonoma de Barcelona, Bellaterra 08193, Spain; email: pablo.vales@icn2.cat

[2] Institute of Physics, University of Silesia in Katowice, ul. Uniwersytecka 4, 40-00 Katowice, Poland.

[3] State Key Laboratory of New Ceramics and Fine Processing, School of Materials Science and Engineering, Tsinghua University, Beijing 100084, China.

[4] Institut Català de Recerca i Estudis Avanc̦ats (ICREA), Barcelona 08010, Catalonia; email: gustau.catalan@icn2.cat

**ABSTRACT**

Flexoelectricity (coupling between polarization and strain gradients) is a property of all dielectric materials that has been theoretically known for decades, but it is only relatively recently that it has begun to attract experimental attention. As a consequence, there are still entire families of materials whose flexoelectric performance is unknown. This is the case, for example, of antiferroelectrics: materials with an antipolar but switchable arrangement of dipoles. And yet, these materials could be flexoelectrically relevant because it has been hypothesised that the origin of their antiferroelectricity might be flexoelectric. In this work, we have measured the flexoelectricity of two different antiferroelectrics ($PbZrO_3$ and $AgNbO_3$) as a function of temperature, up to and beyond their Curie temperature. Neither flexoelectricity nor the flexocoupling coefficients are anomalously high, but the flexocoupling shows a sharp peak at the antiferroelectric phase transition.

Antiferroelectricity was first proposed by Kittel in 1951 in a theory based on antiparallel dipolar displacements analogous to the antiferromagnetism picture [1], and it was experimentally reported for the first time at the end of the same year [2] . Compared to their ferroelectric counterparts, however, antiferroelectrics have been less researched, partly due to their relative rarity, but also because, not being polar, their practical applications are less obvious. So far, they have been studied mostly in the context of electrostatic energy storage [3], [4], but also in electrocaloric applications due to their anomalous (negative) effect [5], [6], and for high-strain actuators [7], [8]. Recently, a record-breaking photovoltaic field (6MV/cm, the highest ever measured for any material) has also been reported in $PbZrO_3$, opening a new line for antiferroelectrics in photovoltaic applications [9].

Owing to their centrosymmetric ground state, antiferroelectrics (AFEs) are not suitable for direct piezoelectric transduction (conversion of strain into voltage). They can, however, be flexoelectric (conversion of strain gradient into voltage). This effect is allowed by all crystal symmetries [10] and it is the result of a linear coupling between a strain gradient and polarization that follows the equation:

$$P_i = \mu_{klij} \frac{\partial u_{kl}}{\partial x_j} \tag{1}$$

Mashkevich & Tolpygo [11], [12] were the first ones to propose such an effect, and Kogan [13] later developed the phenomenological theory. Although it was initially predicted that flexoelectricity would be low in simple dielectrics ($\mu \approx 10^{-10}$ C/m), its proportionality to the permittivity [14], [15] meant that it could reach much higher values, of the order of nC/m and even $\mu$C/m in ferroelectrics and relaxors [16]. Moreover, thanks to barrier-layer effects, even bigger effective coefficients (mC/m) can be reached in semiconductors [17]. In addition, flexoelectricity has become a growing field in the last decades with the development of nanoscience, owing to the inverse proportionality between a device's size and the strain gradients that it can withstand [18].

In the case of AFEs, there is specific interest in their flexoelectricity because it was theoretically predicted by Tagantsev *et al.* [19] and also discussed by Borisevich *et al*. [20] that flexoelectric coupling could be responsible for stabilizing the AFE phase. The concept behind such theories is that antiferroelectric ordering can be viewed as a form of polarization gradient, since polarization alternates every other unit cell. If so, the existence of such spontaneous polarization gradients (antipolar arrangements) would suggest the existence of a strong flexocoupling to the lattice mode responsible for the paraelectric to antiferroelectric phase transition [21]. It is the purpose of this paper to examine whether antiferroelectrics display anomalous flexoelectricity by measuring the flexoelectric and flexocoupling coefficients of the archetype AFE material, PbZrO$_3$, and also of pure AgNbO$_3$ a lead-free AFE. [1]

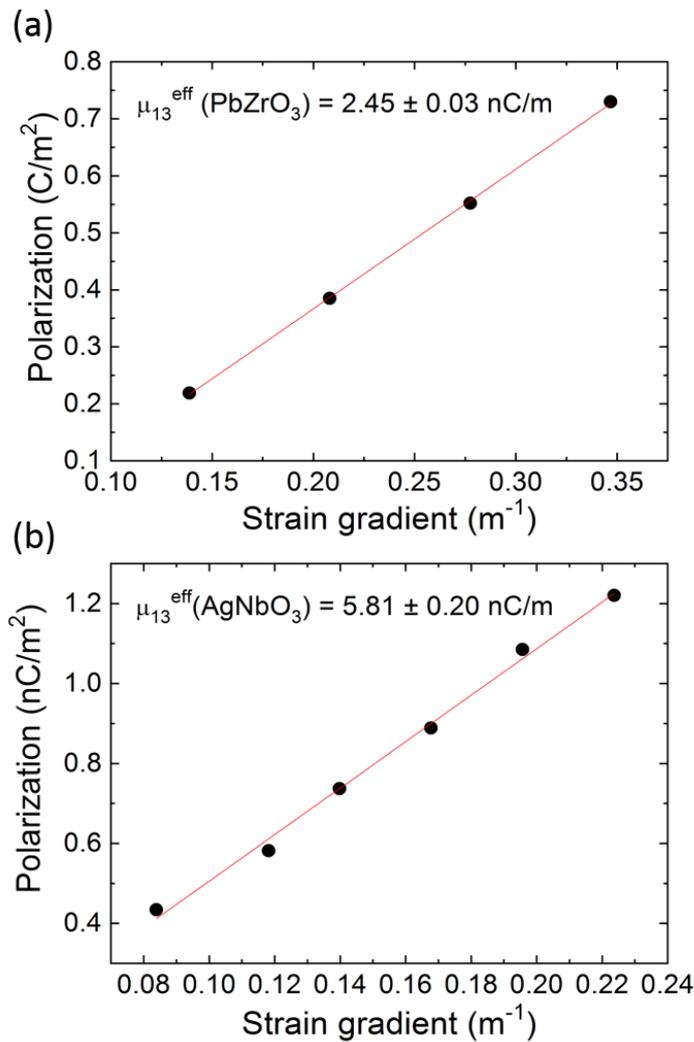

**Figure 1** Measurement of the flexoelectric coefficients of (a) $PbZrO_3$ and (b) $AgNbO_3$ at room temperature. The flexoelectric coefficient is calculated as the slope of the linear fit to the polarization vs strain gradient.

**RESULTS AND DISCUSSION**

Fabrication details and antiferroelectric loops of the ceramic PbZrO$_3$ and AgNbO$_3$ samples are provided in refs. [22] and [3], respectively. Their flexoelectricity has been measured by the method developed by Zubko et al. [23]: a dynamic mechanical analyzer (DMA 8000, Perkin-Elmer) is used to apply a periodic three-point bending stress whilst simultaneously recording the elastic response (storage modulus and elastic loss). The DMA's mechanical force signal is fed into the reference channel of a lock-in amplifier (Stanford Research Instruments, model 830), while the samples' electrodes are connected to the measurement channel of the lock-in amplifier, which records the bending-induced displacement currents. The displacement current is converted into polarization using $P_i = I/2\pi\nu A$, where $\nu$ is the frequency of the applied force (13 Hz in our experiments) and A is the area of the electrodes. The polarization measured by the lock-in is related to the effective flexoelectric coefficient $\mu_{13}^{eff}$:

$$\bar{P}_3 = \mu_{13}^{eff} \frac{\overline{\partial u_{11}}}{\partial x_3} \qquad (2)$$

$$\frac{\overline{\partial u_{11}}}{\partial x_3} = \frac{12z_0}{L^3}(L - a) \qquad (3)$$

where L is the separation between the standing points of the ceramic, *a* is the half-length of the electrodes, and z$_0$ is the displacement applied in the middle of the sample. The mechanical, flexoelectric and dielectric properties were recorded first at room temperature and then as a function of temperature up to 250 °C for the PbZrO₃ and 400 °C for the AgNbO₃. The temperature ramp was 3 °C/min in both cases.

---

[1] Although in AgNbO$_3$, a weak ferroelectric-like polarisation of the order of 4x10$^{-4}$ C/m$^2$ has been reported [31], this residual polarization is thought to be metastable, with the ground state being antiferroelectric [32].

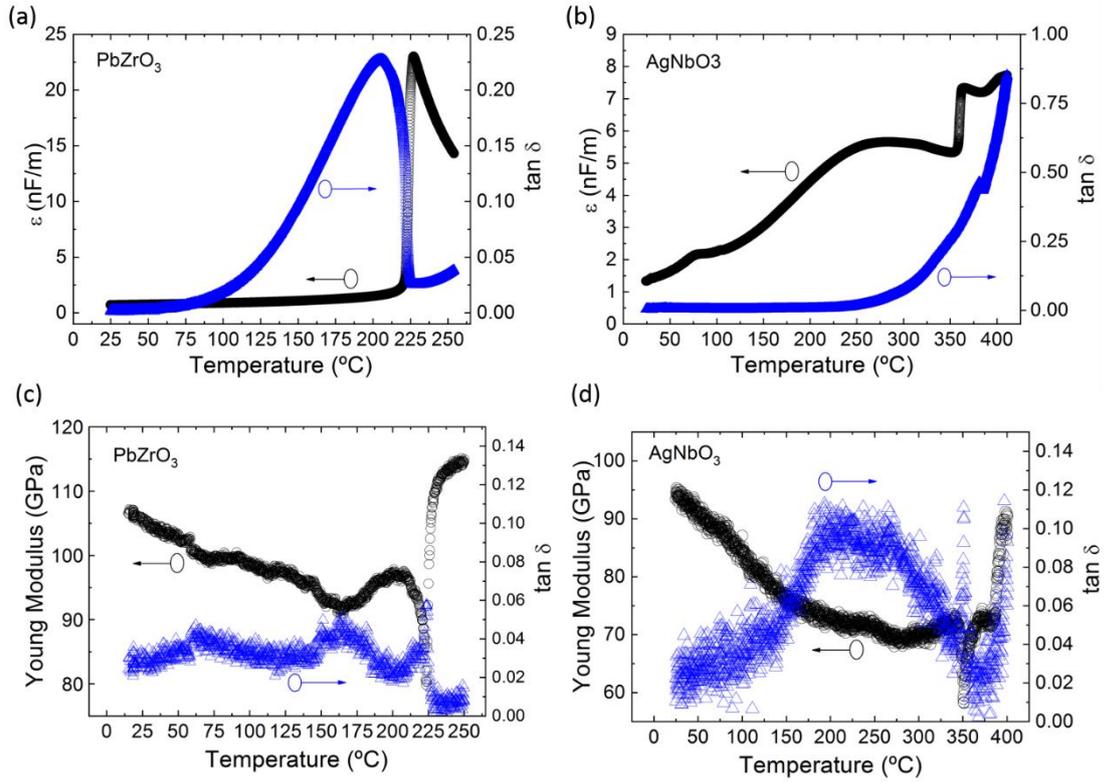

**Figure 2** Permittivity and mechanical properties of (a), (c) PbZrO₃ and (b), (d) AgNbO₃ with their respective phase changes

The room-temperature effective flexoelectricity is shown in Figure (1), where the slope of the linear fit to the data using eq. (2) represents the flexoelectric coefficient. The room-temperature flexoelectric coefficients are $2.45 \pm 0.03$ nC/m and $5.81 \pm 0.20$ nC/m for PbZrO₃ and AgNbO₃, respectively. These room-temperature flexoelectric coefficients are not particularly large; they are considerably smaller than reported for ferroelectrics and relaxors [16], and comparable to the flexoelectricity of SrTiO₃ [23], a non-polar perovskite.

We also calculated the flexocoupling coefficient (flexoelectricity divided by dielectric permittivity), obtaining values of 3.18 and 4.44 V for PbZrO₃ and AgNbO₃, respectively. These values are inside the standard range (1-10 V) predicted [13], [24] and measured [25] for non-antiferroelectric materials, thus not showing the enhancement that might have been expected if antiferroelectricity is driven by flexoelectricity.

On the other hand, room temperature is far below the phase transition temperature of these materials. If flexoelectricity truly has an influence on antiferroelectricity, such coupling should manifest itself most strongly at the phase transition. We therefore characterized the two antiferroelectrics as a function of temperature across their phase transitions. The dielectric and mechanical properties are shown in Figure (2), and the flexoelectric and flexocupling coefficients are shown in Figure (3). Lead zirconate displays a simple Curie-Weiss behaviour as a function of temperature, with a permittivity peak at the critical temperature ($T_C$=225 °C) of the antiferroelectric-to-paraelectric phase transition. Concomitant with this peak, there is an abrupt change (a softening) of the mechanical properties and a maximum in the flexoelectric coefficient, $\mu_{13}^{eff}$. The *flexocoupling* coefficient as a function of temperature, $f_{13}^{eff}$, shown in Figure (3a), stays remarkably constant around 2-3V until, about 50 degrees below $T_C$, it starts to rise, reaching a peak value of 20 V at the transition. Just above the transition, the flexocoupling sharply drops to a value smaller than 1V.

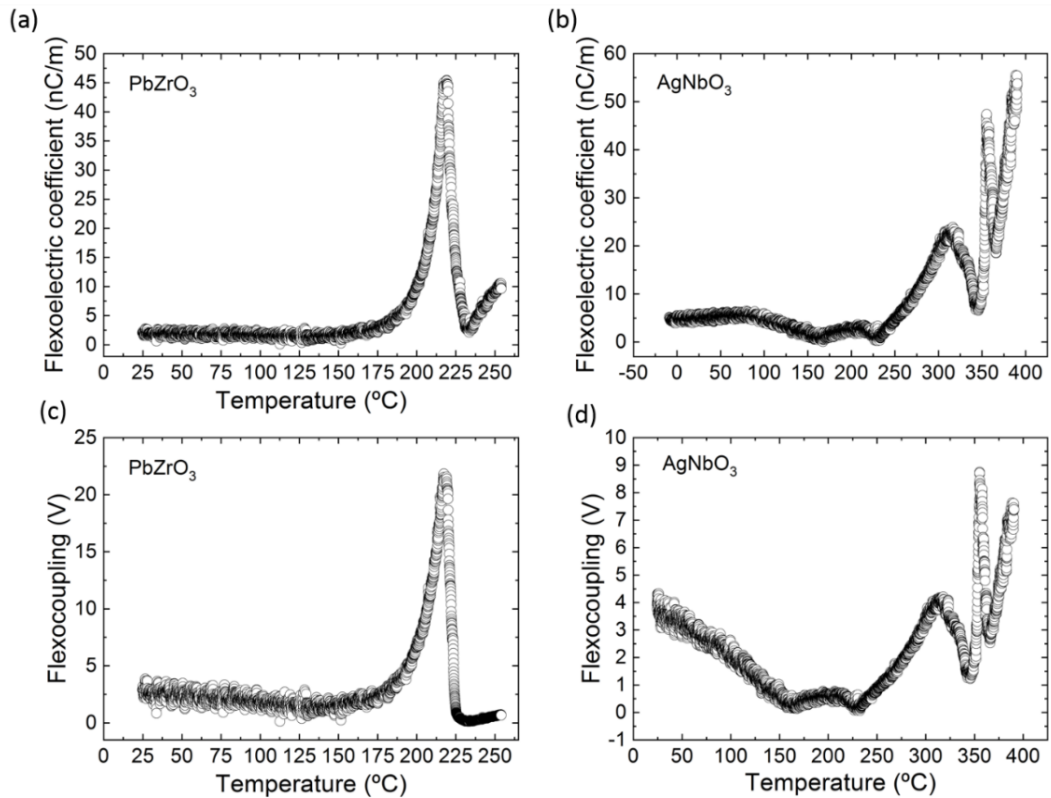

**Figure 3** Flexoelectric coefficient and flexocoupling for (a), (c)  PbZrO₃ and (b), (d) AgNbO₃ up to and beyond their antiferroelectric-to-paraelectric phase transition as a function of temperature

Silver niobate is somewhat more complex, because it has several structural transitions [26] before the antiferroelectric-paraelectric phase transition at $350^{\circ}$C. These phase transitions have a noticeable impact on the flexoelectric coefficient, which shows discontinuities at each of these phase changes, before rising from few nC/m at room temperature to tens of nC/m at the antiferroelectric Curie temperature. The effective flexoelectric coefficient of $AgNbO_3$ continues to rise beyond the Curie temperature, but the dielectric losses also shoot up, suggesting that the high-temperature enhancement in effective flexoelectricity may be due to the increased conductivity [17]. Like the flexoelectric coefficient, the flexocoupling coefficient of $AgNbO_3$ as a function of temperature (Figure (3)) also shows anomalies at all the phase transitions, but in all cases it stays within the moderate range predicted for simple dielectrics (f < 10 V).

The flexoelectriticy of AFE ceramics is therefore not anomalously high. One possible objection to these experimental results is that, below $T_c$, $PbZrO_3$ and $AgNbO_3$ are ferroelastic, and therefore twinning might in principle accommodate part of the induced strain gradient, thus reducing the apparent flexoelectricity coefficient (as has been observed also in $SrTiO_3$ below its ferroelastic phase transition [23]). However, above $T_c$ there is no ferroelasticity, and yet the measured flexocoupling coefficient remains mediocre. Another question concerns the role of surface piezoelectricity, particularly in a ceramic in which grain boundaries provide additional surfaces. However, for the few materials for which we can compare single crystals and ceramics [25], grain boundaries appear to increase, rather than decrease, the effective flexoelectricity. Our conclusion thus remains that the flexoelectricity of antiferroelectrics is not anomalously high. Similar perovskite oxides, such as $SrTiO_3$, have even higher flexoelectric coefficients but do not develop antiferroelectricity, so it is hard to argue that antiferroelectricity is caused by flexoelectricity –at any rate, it is not caused by an anomalously large flexocoupling. This result will have to be taken into account by any theory of the interplay between flexoelectricity and antiferroelectricity [19], [20].

On the other hand, after dividing the flexoelectric coefficient by the permittivity, the resulting flexocoupling coefficient $f$ would be expected to be constant for ordinary materials, bbecause the temperature dependence is mostly contained in the permittivity. In contrast, however, we have seen that the flexocoupling coefficients of our antiferroelectric samples increase sharply near the antiferroelectric phase transition. While their magnitude still remains within the theoretically moderate range, this sharp peak in flexocoupling near $T_C$ is unexplained. In $PbZrO_3$, perhaps part of this increase in effective flexoelectricity could be attributed to the appearance of an intermediate polar phase reported to exist for a few degrees right under the transition [27] and attributed to local strains due to defects in lead and oxygen sublattices [28] combined with strongly anharmonic optic–acoustic mode coupling [29]. However, the observed temperature range of stability of this polar phase [30] is narrower than the width of the observed peak in flexoelectricity. In addition, while polar regions may contribute to the flexoelectric enhancement of $PbZrO_3$, $AgNbO_3$ remains strictly non-polar in temperatures above 75$^o$C, so its flexoelectric peak cannot be associated with parasitic piezoelelectricity. The involvement of flexoelectricity at the critical point of the antiferroelectric phase transition thus appears to be non-trivial and deserves further scrutiny.

**ACKNOWLEDGEMENTS**

We acknowledge financial support from the following grants: Plan Nacional (MINECO Grant MAT2016-77100-C2-1-P), ERC Starting Grant 308023, Severo Ochoa (MINECO Grant no. SEV-2013-0295), MINECO Grant BES-2016-077392, the National Science Centre of Poland (Project 2016/21/B/ST3/02242) as well as the Ministry of Science and Technology of China (Grant no. 2015CB654605).

The author P.V would like to thank the collaboration to F. Vásquez-Sancho for technical assistance with the DMA as well as constructive discussions on the topic.